# Ultrafast switching of a nanomagnet by a combined out-of-plane and in-plane polarized spin-current pulse

O. J. Lee, V. S. Pribiag, P. M. Braganca, P. G. Gowtham, D. C. Ralph and R. A. Buhrman

Cornell University, Ithaca, NY 14853-2501

## ABSTRACT

We report on spin valve devices that incorporate both an out-of-plane polarizer (OPP) to quickly excite spin torque (ST) switching and an in-plane polarizer/analyzer (IPP). For pulses < 200 ps we observe reliable precessional switching due largely to ST from the OPP. Compared to a conventional spin valve, for a given current in the short pulse regime the addition of the OPP can decrease the pulse width necessary for switching by a factor of 10 or more. The influence of the IPP is most obvious at longer, smaller pulses, but also has beneficial ST consequences for short pulse switching.

The spin torque (ST) induced in a ferromagnetic element by a spin polarized current may enable the development of ST magnetic random access memory (ST-MRAM).[1,2] For many applications, ST memory elements should be capable of fast switching, at or below the ns time scale. Fast pulsed-current reversal experiments have previously been performed[3,4,5,6,7,8] on current-perpendicular-to-the-plane (CPP) spin valve devices in which both the polarizing fixed magnetic layer and the switchable free magnetic layer have moments that lie in the sample plane in equilibrium (see Fig. 1b inset). In this conventional geometry, generally there is an incubation time prior to reversal during which stochastically-initiated free layer oscillations grow gradually under ST, and the sub-ns switching process is usually unreliable, with switching probabilities $P_S$ < 1 at the currents of interest for applications, due to thermal fluctuations in the initial magnetic orientation. Reliable switching with sub-ns pulses has been achieved in the conventional structure by adding a hard axis field[9,10] to establish an equilibrium offset angle between the reference and free layers that is ≠ 0 or $\pi$, although this approach adds circuit complexity.

A device modification for achieving fast ST-driven reversal has been suggested by Kent et al.[11] The proposed device has an in-plane polarized free layer (FL) and two fixed magnetic layers: one out-of-plane polarizer (OPP) in addition to one conventional in-plane polarizer/analyzer (IPP) (see Fig. 1d inset). The spin current generated by the OPP exerts a torque on the free layer magnetization tilting it out-of-plane, inducing an out-of-plane demagnetization field that when sufficiently large can quickly rotate the free layer moment to the reversed orientation by a process similar to precessional reversal driven by hard axis magnetic field pulses.[12]

Here we report the fast ST pulse (100 ps - 10 ns) switching performance of devices that incorporate such an OPP layer. These devices are of the type that has been previous employed in spin valve experiments devices that have examined thermally activated switching[13] and microwave emission[14,15,16]. We find that inclusion of the OPP allows ST switching of a nanomagnet to be achieved using simple spin current pulses with pulse width ($t_p$) as short as 100 ps. We demonstrate reliable switching at room temperature provided that $t_p$ is shorter than a critical threshold and the pulse amplitude ($I_p$) is within a relatively broad window (~4 mA). For sub-ns switching, the $I_p$ required for devices with the OPP is much less than for devices with just an IPP fixed layer. We also find that the ST from the IPP, previously assumed[11] to be negligible, has a significant influence on the short-pulse reversal.

We used sputter deposition and e-beam lithography to fabricate CPP spin valve devices with an elliptical cross-section of ~ 70 × 180 nm$^2$ using two different layer structures. The first type had a conventional spin-valve (CSV) configuration consisting of bottom-lead-1/Py(5)/Cu(12)/Py(20)/top-lead (thicknesses in nm), where Py is $Ni_{80}Fe_{20}$, the bottom-lead-1 is Py(5)/Cu(120) and top-lead is Cu(2)/Pt(30). The second type (OPSV) had the additional OPP layer. The layer configuration was bottom-lead-2/OPP/Cu(6)/Py(5)/Cu(12)/Py(20)/top-lead, where bottom-lead-2 is [Ta(5)/Cu(N)(20)]$_2$/Ta(25) and the OPP was Pt(10)/[Co(0.44)/Pt(0.68)]$_4$/Co(0.66)/Cu(0.3)/Co(0.66) (see Fig. 1d inset).[14,17] In both device types the 5 nm Py layer served as the magnetic FL and the 20 nm Py layer was the IPP. The out-of-plane anisotropy for an unpatterned film of the OPP layer was ~7 kOe. All of the ST measurements we report were performed at ~ 300 K under an applied field canceling the average in-plane component of the dipole field from the IPP. Four CSV devices and five OPSV devices were studied in detail and similar results were obtained for all devices of each type.

The average resistance difference $\Delta R$ between the parallel (P) and anti-parallel (AP) configurations of the CSV devices was $110 \pm 15$ m$\Omega$, while for the OPSV devices $\Delta R = 85 \pm 5$ m$\Omega$. This difference may be due to spin scattering in the OPP and/or to the effect of the dipole field from the OPP, which acts to cant the FL moment slightly out of plane.

We first measured the average currents for thermally-activated switching of the free layer, both from AP to P (AP-P, relative to the IPP) and from P to AP (P-AP) as the function of the current ramp-rate to determine[18,19] the energy barrier ($E_a$) for magnetic reversal and the zero-thermal-fluctuation critical current ($I_{c0}$). For a representative pair of devices we obtained $I_{c0}^{AP-P} \sim -2.65$ mA, $I_{c0}^{P-AP} \sim 2.58$ mA, $E_a^{AP-P} \sim 1.74$ eV, and $E_a^{P-AP} \sim 1.79$ eV for the CSV device, and $I_{c0}^{AP-P} \sim -2.42$ mA, $I_{c0}^{P-AP} \sim 2.50$ mA, $E_a^{AP-P} \sim 1.40$ eV, and $E_a^{P-AP} \sim 1.42$ eV for the OPSV device. We attribute the somewhat lower values of $E_a$ in the latter case to the effect of the dipole field from the OPP in decreasing the effective in-plane anisotropy field ($H_k^{\text{eff}}$).

Results of pulsed-current ST reversals are shown in Fig. 1a-d, which plots the switching probability ($P_s$) for quasi-rectangular (~65 ps rise and 105 ps fall time) pulses as a function of $I_p$ and $t_p$. The CSV devices show reliable switching by the 6 ns (full width at half maximum) pulses, but 100% switching probability is impossible with 100 ps pulses up to $|I_p| \sim 16$ mA (Fig. 1a-b). The OPSV devices exhibit three regimes of behavior in Fig. 1c-d: (i) a long pulse regime (e.g., $t_p$ = 6 ns), where the switching distributions of the OPSV are very similar to the CSV, up to a certain $|I_p|$; (ii) an intermediate pulse-width regime (e.g., $t_p$ = 600 ps), where there is no reliable switching of the OPSV; and (iii) a short pulse regime, $t_p \leq 0.2$ ns, where the OPSV reversal is very reliable (more than 998 reversals in 1000 attempts over a significant range of $I_p$) and efficient, with a much lower $I_p$ required for switching compared to the CSV. For $t_p$ = 100 ps, at very high currents, ~ twice the onset current for first achieving $P_s$ = 100%, $P_s$ begins to decrease, which we attribute to over-rotation in the precessional reversal. For 0.2 ns < $t_p$ < 1 ns this over-rotation due to the OPP ST makes it impossible to obtain reliable OPSV reversal, while for long pulses, > 1 ns, the additional OPP ST results in only a limited range of pulse amplitude where 100% reversal can be obtained.

In Fig. 2 we plot the values of $I_p$ that yielded $P_s$ = 95% for the OPSV and CSV as the function of $1/t_p$ to compare the ST-induced switching speeds. In the macrospin approximation for $I > I_{c0}$ the switching time $\tau$ for a CSV varies linearly with ST current amplitude[3] as $\tau^{-1} = \zeta(I - I_{c0})$. Fitting to the CSV data of Fig. 2, assuming that $\tau \approx t_p$, we obtain $\zeta^{AP-P} = 0.158$ ns$^{-1}$mA$^{-1}$, $I_{c0}^{AP-P} = -2.55$ mA and $\zeta^{P-AP} = 0.131$ ns$^{-1}$mA$^{-1}$, $I_{c0}^{P-AP} = 2.44$ mA. These $I_{c0}$ values are in close accord with the values obtained from the ramp-rate measurements for the CSV. The same linear relationship also provides a good fit for the OPSV switching data in the short pulse regime ($1/t_p > 5$ ns$^{-1}$) despite the fact that the assumptions of ref. 3 do not apply. Fits to the OPSV data in Fig. 2 yield $\zeta^{AP-P} = 6.117$ ns$^{-1}$mA$^{-1}$, $I_{c0}^{AP-P} = -5.13$ mA, and $\zeta^{P-AP} = 11.54$ ns$^{-1}$mA$^{-1}$, $I_{c0}^{P-AP} = 3.54$ mA. These $I_{co}$ values are significantly larger than those obtained from the ramp-rate data, suggesting that the OPSV reversal mechanism for short $t_p$ is distinctly different than for long $t_p$. Moreover, the short-pulse ST switching speed efficiency ($\zeta$) is approximately $40\times$ that of the CSV device for the AP-P case, and nearly $90\times$ that for the P-AP case. We ascribe this to the lack of an incubation delay in the OPSV, reflecting that precessional reversal in the OPSV need not be preceded by a slow spiraling of the FL moment away from the equilibrium configuration.

The difference in the mechanisms for fast-pulse switching in the OPSVs and CSVs is also illustrated by $\Delta I_p$, the difference between the pulse amplitudes required for 20% and 80% switching probabilities (see Fig. 2 insets). For the CSV, $\Delta I_p$ grows to be as large as 5 mA, while for the OPSVs in the short pulse regime $\Delta I_p$ is always less than 0.7 mA. ($\Delta I_p$ for OPSVs can be larger for longer pulses; see Fig. 2. insets) The broad distributions for the CSVs can be explained by thermal fluctuations in the initial offset angle of the free layer about the P and AP configurations. Because the initial orientation of the FL in the OPSV is always close to perpendicular to the OPP, the effects of thermal fluctuations are minimized.

The original proposal of Kent et al. for OPP precessional reversal anticipated that a current pulse of either bias would equally well drive magnetic reversal for either P-AP or AP-P switching.[11] This current symmetry would limit the write operation to toggle mode (or reversible) switching where the final state is always flipped from the initial state. However we observe that the minimum values of $|I_p|$ required for short-pulse reversal are different for AP-P and P-AP switching (Fig. 1c-d), and are also different when the current flows are reversed (Fig. 1e-f). For the OPSVs, P-AP switching requires lower onset currents than AP-P, and switching is also easier for the sign of $I_p$ that gives ST-switching in the CSVs (Fig 1c-d) than for reversed currents (Fig 1e-f).

Based on micromagnetic simulations, we argue that the differences in onset current between P-AP and AP-P reversals are due to the combined effect of the dipolar fields from the edges of the IPP and from the OPP, which add on one side of the free layer (the right side in the insets of Fig. 1(e,f)) but almost cancel on the other. This non-uniform field causes the effective in-plane anisotropy field $H_k^{eff}$ on the additive side (right) to increase in the AP case and decrease in the P case, giving effectively different onset currents[20] for reversal as a function of position. The result in simulations is that reversal first occurs at one end of the FL (on the right for P-AP, and the left for AP-P, independent of the sign of $I_p$) and then is completed via the exchange interaction, and that the value of $|I_p|$ needed for reversal is lower in the P-AP case than for AP-P.

The effect of ST from the IPP can explain the difference in short pulse reversal behavior with current direction for a given type of switching (P-AP or AP-P), in that just as in CSV devices the ST from the IPP promotes P-AP switching for $+I_p$ and AP-P switching for $-I_p$. Due to the greater non-uniformity in the starting magnetization in the AP configuration, the effect of the IPP ST is enhanced, giving a larger difference between the two signs of current for AP-P reversal (compare Fig. 1f and 1c). These differences provide a current window [$\Delta(+I_p) \approx 6$ mA] in which a $+I_p$ can reliably drive P-AP switching without AP-P. This can alleviate the need to employ a read-before-write approach in short pulse OPP ST-MRAM devices that would be required if the threshold values of $I_p$ were equal.

In summary, we show that reliable precessional switching with short ($t_p < 0.2$ ns) rectangular pulses can be achieved in ST devices incorporating both OPP and IPP fixed layers. Due to the effects of the IPP ST and non-uniform local dipole fields, we find different threshold currents for the four cases $+I_p^{P\text{-}AP}$, $-I_p^{P\text{-}AP}$, $+I_p^{AP\text{-}P}$, and $-I_p^{AP\text{-}P}$. The results indicate that it is possible to optimize pulse amplitudes and widths within significant parameter windows so that a pulse with a given sign of current produces only the desired state (P or AP). Such devices could lead to a very high-speed non-volatile magnetic memory cell with sub-100 ps write pulses.

This work was supported by the National Science Foundation (NSF)/Nanoscale Science and Engineering Center program through the Cornell Center for Nanoscale Systems, by the Office of Naval Research, and by the Semiconductor Research Corporation, and was performed in part at the Cornell NanoScale Facility, a node of the National Nanotechnology Infrastructure Network (NSF Grant No. ECS 03-35765) and in the facilities of the Cornell Center for Materials Research, which is supported by the NSF/Materials Research Science and Engineering Center program.

FIGURE CAPTIONS

Fig. 1 (color online): Switching probability $P_s$ as the function of pulse amplitude $I_p$ for 100 ps, 600 ps and 6 ns pulse widths ($t_p$). (a,b) Results for the CSV for (a) AP-P and (b) P-AP reversal. Inset: Schematic of the CSV device, (c,d) Results for the OPSV for the signs of current which give switching in conventional ST devices: (c) AP-P and (d) P-AP. Inset: Schematic of the OPSV device. (e,f) Results for the OPSV with the signs of current opposite to those needed for switching in conventional ST devices: (e) P-AP and (f) AP-P. Inset: Micromagnetic simulation of the configurations for the initial P and AP states.

Fig. 2 (color online): Comparison of the reversal speed between the OPSV and CSV devices. The inverse of pulse widths ($1/t_p$) is plotted as the function of the pulse amplitude $I_p$ that yields a 95% switching probability $P_s$ for: (a) P to AP and (b) AP to P reversals. For the OPSV device, pulse widths between 0.3 ns and 2 ns do not achieve 95% AP-P switching for any negative value of $I_p$. Insets: The difference ($\Delta I_p$) between the pulse current amplitudes that yield $P_s = 80\%$ and $P_s = 20\%$ as a function of $t_p$ for (a) P-AP and (b) AP-P. Large values of $\Delta I_p$ indicate a significant effect of thermal fluctuations on the reversal process.

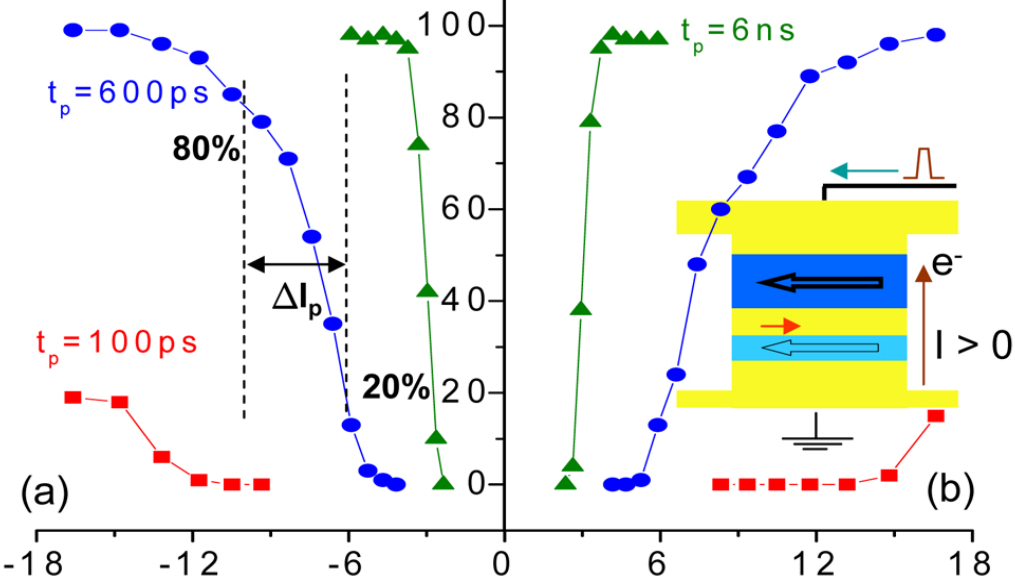

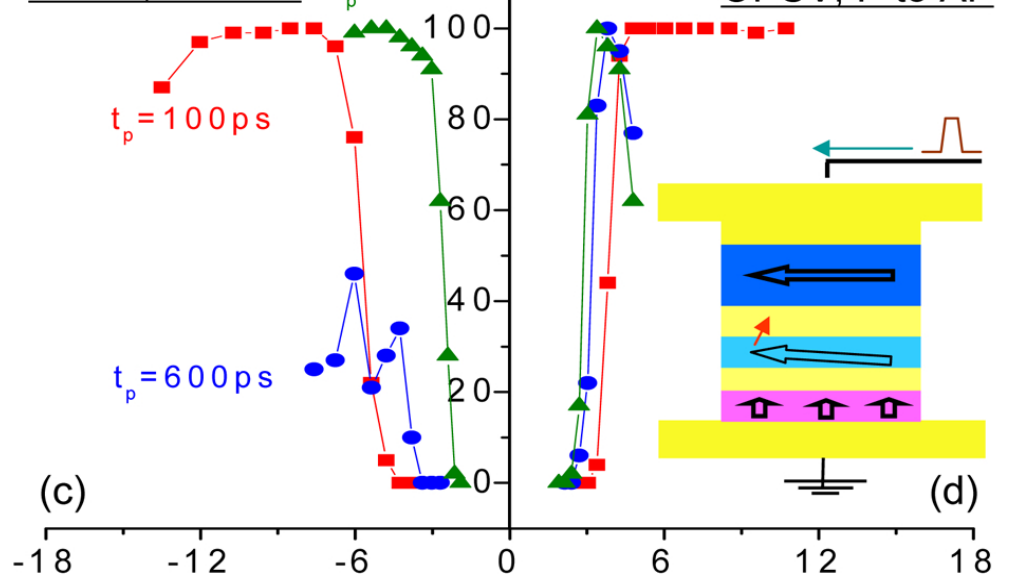

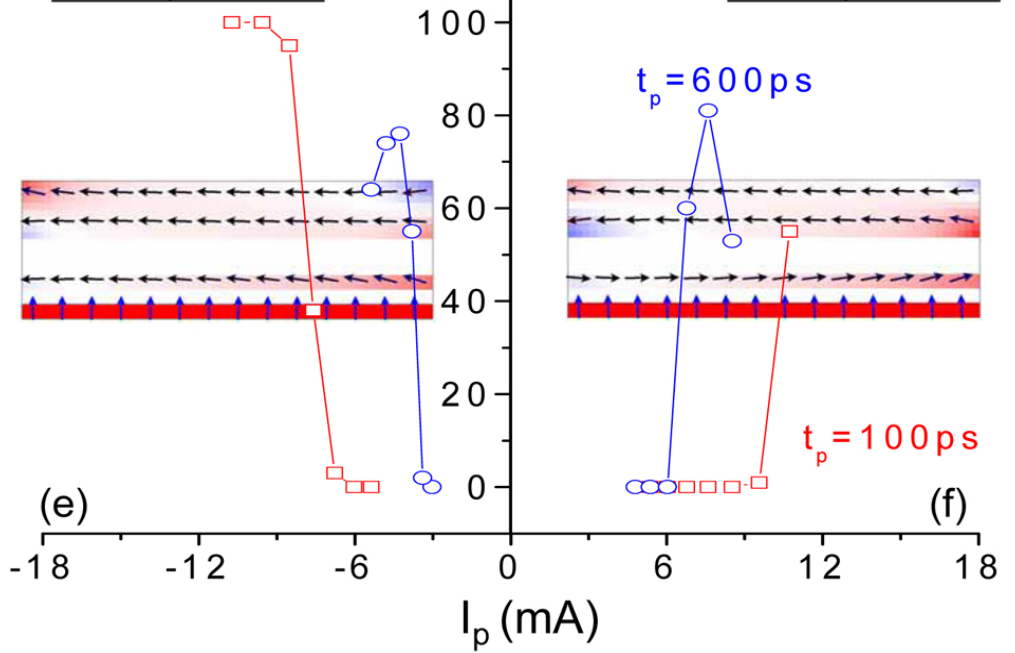

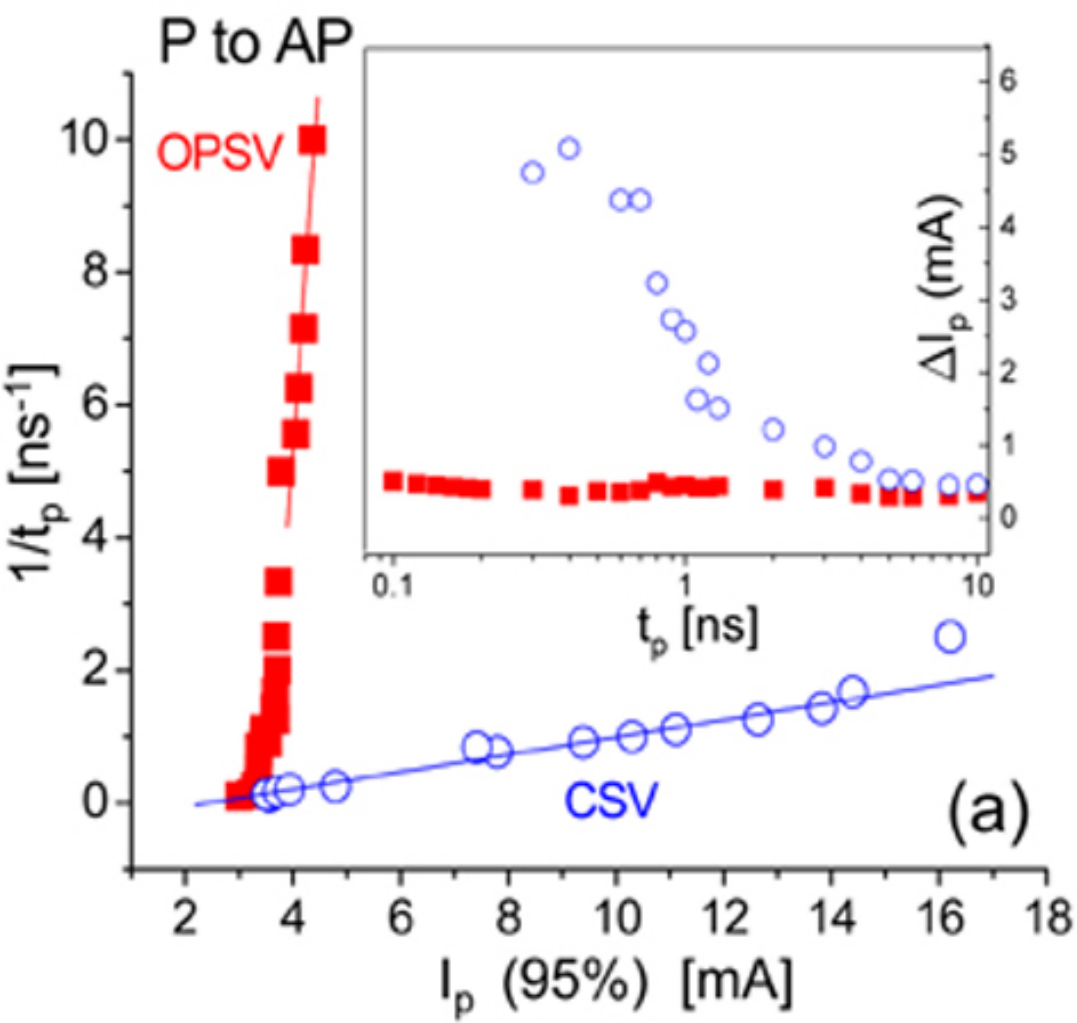

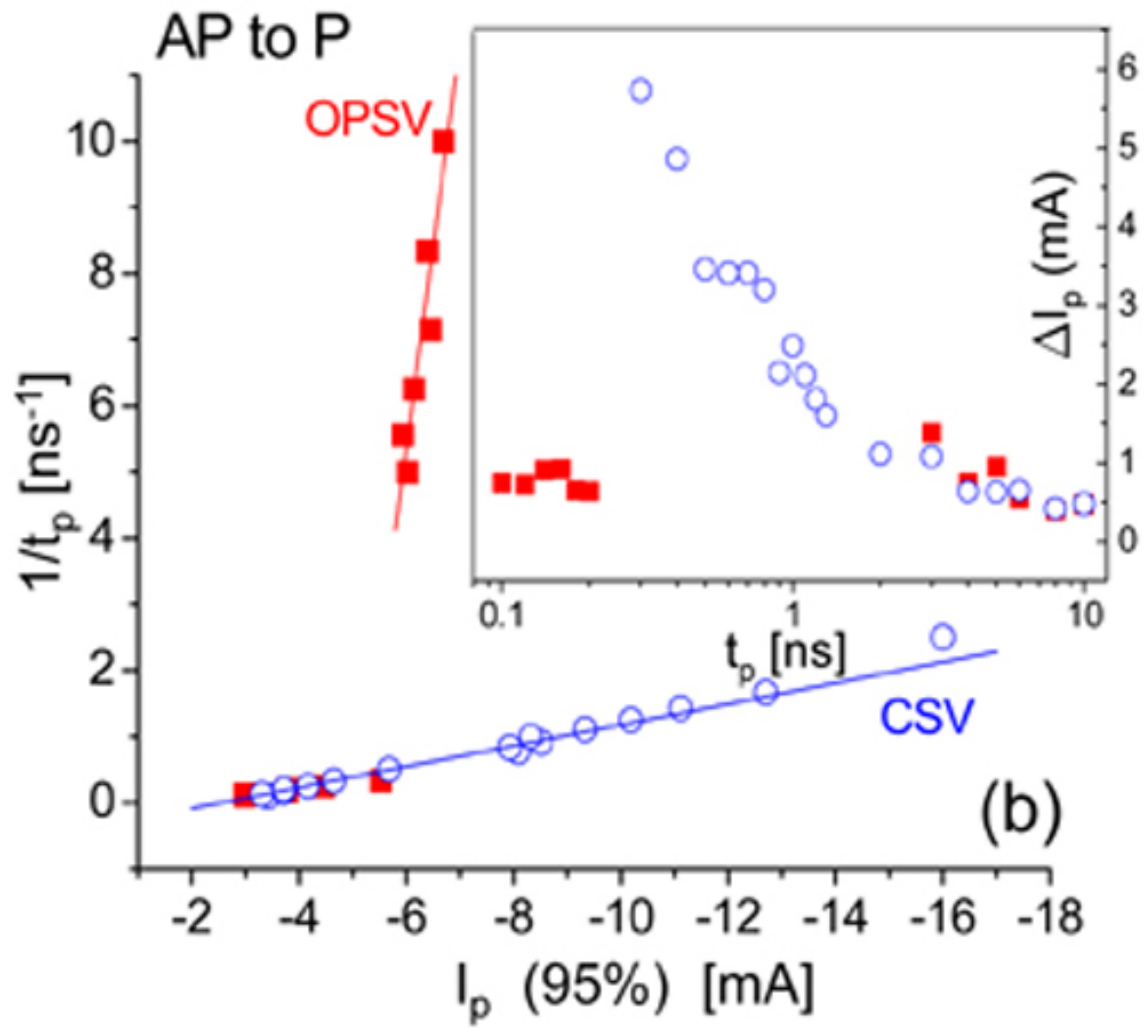